\documentclass[reprint, amsmath,amssymb,aps, prl]{revtex4-1}
\usepackage[utf8]{inputenc}
\usepackage[T2A]{fontenc}
\usepackage{indentfirst}
\usepackage{graphicx}
\usepackage{amsmath}
\usepackage{verbatim}
\usepackage{color}
\renewcommand{\Re}{\mathop{\rm Re}\nolimits}
\renewcommand{\Im}{\mathop{\rm Im}\nolimits}

\begin{document}
\date{\today}
\title{Temperature gradient and Fourier's law in gradient-mass harmonic systems} 


\author{K.V. Reich}
 \altaffiliation{Ioffe Physical-Technical Institute, 194021 St Petersburg, Russia}
\email{Reich@mail.ioffe.ru}

\begin{abstract}
Heat flow and thermal profile in a 1D harmonic lattice with coordinate-dependent masses has been calculated in the thermodynamic limit. It is shown in the particular example of a 1D harmonic lattice with linearly increasing masses that in standard Langevin conditions of contact, a temperature gradient can form, and Fourier's law can be obeyed.
\end{abstract}



\maketitle

\section{Introduction}
Raising the efficiency of heat dissipation in nanodimensional systems is presently a major problem whose solution will define the potential of further miniaturization of electronic devices. This problem is complicated by emergence of a host of effects which are specific of heat transport in nanodimensional systems. To cite an example, heat transport is affected noticeably not only by heat resistance at interfaces playing a dominant part in heat transport in nanostructures \cite{Kapitza} but by variations of the pattern of the heat transport itself \cite{conductance,cahill} , when the phonon mean free path becomes comparable with the sample size. This results in the heat transport becoming anomalous: more specifically, the Fourier law does no longer hold, in other words, the heat flux through the system becomes dependent now not on the gradient but on the temperature difference, a phenomenon which has recently been demonstrated experimentally on nanotubes \cite{break, balandin}.

The pursuit of this goal has been a major motivation for studying not only methods that could reduce the heat resistance at interfaces, with recent progress in this direction been reported in Ref. \cite{Reich}, but ways that could lead to development of thermal rectifiers, i.e., to the possibility of varying the modulus of heat flux by changing the sign of the temperature difference applied to the system \cite{rectification_rev}. Considerations of a general nature seem to lead to an obvious assumption that in order to observe such a ``heat rectification effect'', one should produce asymmetry in a system. A theoretical analysis  \cite{Mohammad, rectification_graded} and an experimental study of a nonuniformly mass-loaded nanotube \cite{Experiment} have demonstrated that one-dimensional structures with increasing masses are possible candidates for realization of the effect.

Obviously enough, development of such gradient-mass structures is technologically anything but a simple problem \cite{Roberts2011648}, and we are witnessing presently only the beginning of this process, with the effect of heat rectification not yet realized in full measure. This is why a search for its realization is being pursued along more than one direction. It was proposed \cite{transmission} to use filaments of doped silicon. In place of a spatially varying mass, an idea was also advanced to subject a nanotube to nonuniform tension \cite{gradons}, which is obviously equivalent to variation of mass \cite{PhysRevB.76.205419, PhysRevLett.95.116803}. One could apparently employ for this purpose nanodiamond-decorated carbon nanotubes \cite{Vul}. A possibility is also being discussed of using asymmetric graphene and silicon structures \cite{nanohorns,silicon}.

It appears a plausible assumption that, similar to the p-n junction which has become a basis of electronics, development of gradient-mass materials will form a foundation for progress along the lines of a new domain - phononics \cite{RevModPhys}. As a weighty argument for this conclusion may serve proposals of a number of nanodevices based on such materials, to wit, heat diode \cite{PhysRevLett.97.124302, PhysRevLett.93.184301}, heat transistor \cite{transistor}, heat logical element \cite{logic}, memory devices \cite{memory}, and heat limiter \cite{limiter}.

The above illustrates an increasing interest to studies of thermal properties of gradient-mass systems. Numerous attempts are being undertaken to investigate such materials by both numerical \cite{Mohammad, transmission,PhysRevB.76.020301,PhysRevB.61.6645} and analytical methods \cite{ThermalRectification3,rectification_graded}. Many theoretical aspects remain, however, open, even without inclusion of anharmonic effects into consideration. To cite an example, it was shown numerically that a temperature gradient forms in systems with linearly \cite{gradons} or exponentially \cite{HeatExponentGraded}  varying masses, even in a harmonic case. This phenomenon is surprising in itself. Indeed, despite numerous attempts, a rigorous analytical microscopic foundation of Fourier’s law is still lacking \cite{review,Heated_debate}. Only for several systems has one managed to obtain an analytical result, more specifically, systems with identical \cite{Firstsolution}, alternating \cite{Heatflow2,PhysRevE.76.031116,Heatflow1,o'connor:692} or random masses \cite{Transmission_coefficient,PhysRevE.78.051112,PhysRevE.76.011111}. It should be stressed, however, that in neither of these cases does the Fourier law hold and temperature gradient does not form. The temperature profile is linear and Fourier's law holds only in effective models such as the harmonic chains with self-consistent stochastic reservoirs at each site \cite{netto_Lebowitz_Lukkarinen_2004, PhysRevE.76.031116,PhysRevE.70.046105}. 

We are going to show below that gradient-mass systems possess truly unique properties; indeed, by now these are the only systems in which one can obtain analytically a temperature gradient and make the Fourier law hold without self-consistent stochastic reservoirs.

Significantly, such functionally graded materials, i.e., nonuniform gradient-mass systems, can be met in natural life \cite{He2013}. Note that the optical properties of systems with graded dielectric permittivity match gradient-mass systems \cite{Opt1,Opt2}  . And it is these systems that are attracting great current interest \cite{PhysRevB.77.104204,wei:074102} as an effective medium for application of the Faraday effect \cite{miao:023512}.





\section{Model}
\begin{figure}
\includegraphics[width=1\linewidth]{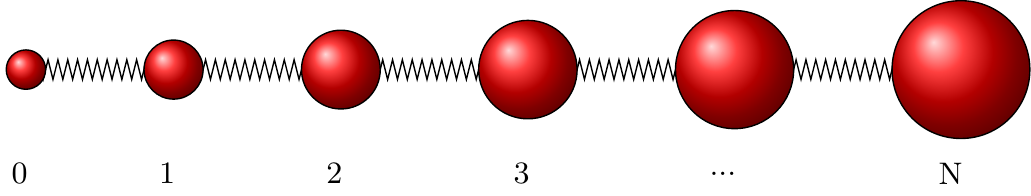}
\caption{(Color  online) Schematic of a harmonic system with coordinate-dependent masses}
\label{fig:chain}
\end{figure}

Consider a one-dimensional chain of $N + 1$ particles which interact harmonically with their nearest neighbors with a spring constant $K$. The momentum and displacement $n$ of a particle of mass $m_n$ will be denoted by $p_n$  and  $x_n$, respectively. The Hamiltonian of this system (Fig. \ref{fig:chain}) can be written in the form:

\begin{equation}
\label{eq:first_hamiltonian}
H= \sum \limits_{n=0}^{N} \left( \dfrac{p_n^2}{2 m_n}+\dfrac{1}{2}K(x_{n+1}-x_{n})^2 \right),
\end{equation}
The equations of motion for such a system become:
\begin{equation}
\label{eq:x}
x_n \cos \beta_n = \dfrac{1}{2}\left( x_{n+1} + x_{n-1} \right),
\end{equation}
where $\cos \beta_n=( 1-{\omega^2 m_n}/{2K}).$ 

In the thermodynamic limit the solution of this system of equations can be obtained in analytical form.

To do this, we use Fourier's method by writing  $x_n$ in the form

\begin{equation}
\label{eq:x_n}
x_n=\int \limits_{k_1}^{k_2} f(k) \exp(ikn) dk.
\end{equation}

We choose the limits of integration $k_1,k_2$ such that the function $f(k)$ and all its derivatives tend to zero at these points. In this case we come to the equality $$(-in)^px_n=\int \limits_{k_1}^{k_2} \frac{d^pf(k)}{dk^p} \exp(ikn) dk.$$ We further assume that the mass $m_n$ and $\beta_n$ are functions of $n/N$. Now the function $\cos \beta(n/N)$ can be expanded in the Taylor series with respect to the variable $n/N$, with Eqs. \eqref{eq:x} converting into a differential equation for function $f$:
\begin{equation}
\label{eq:diff_f}
\cos \beta\left(i\frac{1}{N} \frac{d}{dk}\right)  f  = f \cos(k),
\end{equation}
Only in rare cases, such as a linear dependence of particle masses on index:
\begin{equation}
\label{eq:M}
m_n= M_{0}+\dfrac{M_{1}-M_{0}}{N}n,
\end{equation}
this differential equation allows an exact solution. 

Because to this case we are going to revert more than once, we note that
Eqs. \eqref{eq:x} are solved in terms of Hankel functions: $x_n=H^{(1)}_{n+\Delta n}(z), H^{(2)}_{n+\Delta n}(z)$, where $\Delta n=  z \cos \beta(0), z = N/ (\cos \beta (1) -\cos\beta(0))$. This can be verified by direct substitution using the recurrent relation $2n/zZ_n(z)=Z_{n+1}(z)+Z_{n-1}(z)$,  $Z_n(z)=H^{(1)}_n(z),H^{(2)}_n(z)$ \cite{Bessel_watson}.

In a general case, we are interested in the solution in the thermodynamic limit. We are going to look for the solution to Eq. \eqref{eq:diff_f} subject to the
condition  $N \rightarrow \infty$ in the form: 
$$f(k)=\varphi(k)e^{ i g(k) N}.$$
In this case we come to the following relations for $g$ and $\varphi$:
\begin{equation}
  g=-\int^k \beta^{-1} (z) dz,
\end{equation}
\begin{equation}
\varphi(\beta(k))=\exp\left(- \int^{k} \frac{dx}{M'(x)} \left(\frac{M(x)- M(0)}{x}\right)' \right),
\end{equation}
where $\beta^{-1}$ is a function inverse of $\beta$. Note that $\varphi$ does not depend on frequency.

Next we choose the path of integration in Eq. \eqref{eq:x_n} such that it will descend most steeply from the saddle point. Substituting the expressions for $g$ and
$\varphi$, we come in the limit that $N\rightarrow \infty$ to a particular solution to Eqs. \eqref{eq:x}:

\begin{equation}
\label{eq:y_n}
y_n= \frac{1}{\sqrt{\sin\beta(a)}} \phi(a) e^{ib\beta(a)} e^{iN\int \limits_0^a \beta(x)dx } ,
\end{equation}
where $n$ is taken in the form $n=aN+b$,  $\phi(k) = \varphi(\beta(k)) \sqrt{|M'(k)|}$. 

We finally come to the general solution of Eqs. \eqref{eq:x} in the form $x_n=Ay_n+By_n^*$. 

\section{Flux and the temperature profile}
Using  the standard non-equilibrium Green function method \cite{NEG1, review, transmission} and assuming the temperature on the left to be $T_L=T+1/2 \Delta T$, and on the right, $T_R=T-1/2 \Delta T$, with $\Delta T, \omega_n \ll T$, we can obtain both the heat flux $J$ in the system, and the temperature profile $T_n$:

\begin{eqnarray}
\label{eq:TandJ}
&& J=\frac{\gamma}{m_N}\Delta T I_N, \nonumber \\
&& T_n=T+\frac{1}{2} \Delta T (I_n-\bar{I}_n),
\end{eqnarray}
where the quantities
\begin{eqnarray}
\label{eq:I}
&& I_n=\frac{m_n\gamma}{\pi} \int \limits_{-\infty}^{\infty} d\omega \omega^2 |G_{0n}|^2, \nonumber \\ 
&&\bar{I}_n=\frac{m_n\gamma}{\pi} \int \limits_{-\infty}^{\infty} d\omega \omega^2 |G_{nN}|^2
\end{eqnarray}
are expressed in terms of the Green's function:
$$
G(\omega)=\left(  K - M\omega^2 - \Sigma_L(\omega) - \Sigma_R(\omega) \right)^{-1}.
$$

with $M$ being a diagonal matrix with elements corresponding to the particle masses, and $K$, the dynamic matrix for the system. The function $\Sigma(\omega)$ specifies the conditions of contact of the system under consideration with heat reservoirs. For the standard (Langevin) contact, $\Sigma_L(\omega)=i\omega\gamma \delta_{i,0}\delta_{j,0}$, $\Sigma_R(\omega)=i\omega\gamma \delta_{i,N}\delta_{j,N}$, $\gamma$ is dissipation constant. After a few straightforward transformations similar to those made in Ref. \cite{determinant}, we come to:
\begin{equation}
\label{eq:green_function}
G_{0,k}=\frac{\left|D_{k+1,N}-\Sigma D_{k+1,N-1}\right|^2}{\left|D_{0,N} - \Sigma(\omega) (D_{1,N}+D_{0,N-1}) + \Sigma(\omega)^2 D_{1,N-1}\right|^2},
\end{equation}
where $D_{l,m}$ is defined to be the determinant of the submatrix of $K-\omega^2M$ beginning with the $l$-th row and column and ending with the $m$-th row and column. 
This determinant can be readily derived, because we know the general solution for Eqs. \eqref{eq:x}:
\begin{equation}
  \label{eq:D}
D_{lm}=\frac{\Im y_{l-1}y_{m+1}^*}{\Im y_{l-1}y_l^*}
\end{equation}

\begin{figure}
\includegraphics[width=1\linewidth]{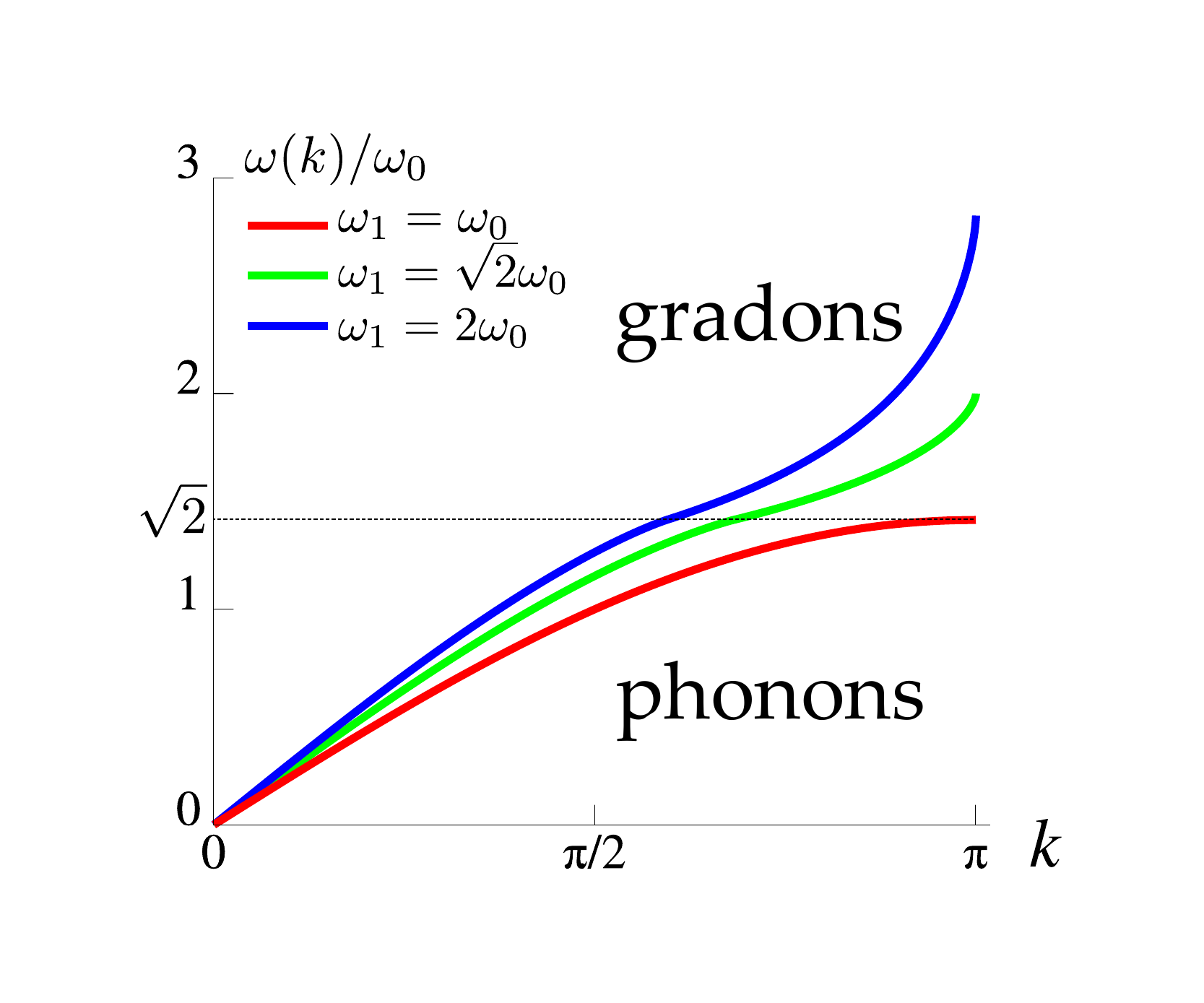}
\caption{
(Color  online) Dispersion relation of the vibration energy in a harmonic chain with linearly increasing masses \eqref{eq:M} plotted vs. the wave vector for different relations between the boundary masses $\omega_{0,1}=\sqrt{2K/M_{0,1}}$. For energies $\omega<\sqrt{2}\omega_0$, the system operates with delocalized phonons, and for $\sqrt{2}\omega_0<\omega<\sqrt{2}\omega_1$, with localized gradons.}
\label{fig:dispersion}
\end{figure}

The solution of equation $D_{0N}=0$ with respect to frequencies $\omega$ yields the dependence of the wave vector $k$ on the frequency of vibrations in the system under consideration $k(\omega)$:
\begin{equation}
  \label{eq:dispersion}
Re \int \limits_0^1 \beta(x) dx =k(\omega)
\end{equation}
In particular, for the case of linear mass distribution \eqref{eq:M} we come to: $$k(\omega)=\Re \frac{f(\beta(0)) - f(\beta(1))}{\Delta_{ \cos}},$$ where $f(\beta)=\left(\sin \beta-\beta\cos \beta \right)$. This dispersion relation is displayed in Fig. \ref{fig:dispersion}. This result correlates with numerical simulations \cite{Gradonproperties}.

As seen from Fig. \ref{fig:dispersion}, in systems with linearly increasing masses one can identify two kinds of vibrations, more specifically, delocalized phonons with frequencies $\omega<\sqrt{2}\omega_0$ , and localized ``gradons'' with frequencies $\sqrt{2} \omega_0<\omega< \sqrt{2} \omega_1$, $\omega_{0,1}=\sqrt{2K/M_{0,1}}$.

In a general case, there exist also phonons with frequencies $\omega<  \sqrt{2} \omega_{min}$ and gradons with frequencies $\omega> \sqrt{2} \omega_{min}$, with $\omega_{min}$ being determined by the maximum mass in the system. Incidentally, for gradons $\beta$ becomes imaginary, with the determinant starting to grow exponentially. As a result, frequencies $\omega> \sqrt{2} \omega_{min}$  do not contribute to the integrals in Eqs. \eqref{eq:I}.

Standard methods of averaging \cite{Heatflow2} permit reducing the integrands in Eq. \eqref{eq:I} in the $N \rightarrow \infty$ limit to the form
\begin{eqnarray}
\label{eq:green}
 && |G_{0N}|^2= \left(\frac{\phi(0)}{\phi(1)}\right)^2  \frac{\sin \beta(0) \sin \beta(1)}{\gamma\omega(1+\gamma^2\omega^2) (\sin \beta(0)+\sin \beta(1))}  \nonumber \\
 && |G_{0n}|^2=\frac{1}{2\gamma\omega} \left(\frac{\phi(0)}{\phi(x)}\right)^2  \frac{\sin\beta(0)}{(\sin\beta(0)+\sin\beta(1)) \sin \beta(a)} \nonumber \\
 && |G_{nN}|^2=\frac{1}{2\gamma\omega} \left(\frac{\phi(x)}{\phi(1)}\right)^2  \frac{\sin\beta(1)}{(\sin\beta(0)+\sin\beta(1)) \sin \beta(a)} \nonumber \\
\end{eqnarray}
where $n$ is used in the form $n = aN$. To preclude misunderstanding, we note that these relations do not go one into another at $a = 1$ and $a = 0$, because
$N \rightarrow \infty$ is a non-uniform limit.

Putting relations \eqref{eq:green} into \eqref{eq:TandJ}, we come to a general answer for the heat flux and temperature profile in a system with an arbitrary mass distribution. While these integrals cannot be performed in a general case, they permit a number of conclusions.

To begin with, it turns out that a flow in a system depends only on the boundary masses and the maximum mass in the chain (which specifies the minimum frequency). We readily see that if the masses are constant, we come to the standard answer for the heat flow \cite{Firstsolution}. In the reverse approximation,  when $M_1 \gg M_{0}$, we obtain:
$$J = \frac{2\sqrt{2} \Delta T}{\pi \omega_{0} \gamma^2} \left(\frac{\phi(0)}{\phi(1)}\right)^2 \left( \sqrt{2}\omega_{min}\gamma - \arctan(\sqrt{2}\omega_{min}\gamma)\right)$$

Up to this point, we assumed the boundary masses not to depend on $N$. We can include this dependence in the case of linearly increasing masses. Setting  $M_1 \sim M_0 N^s$ , the heat flux will acquire the form:
$$J=\frac{8 \Delta T}{3 \pi} \omega_{0}^2 \gamma \frac{1}{N^{3s/2}}.$$
It is easy to verify the correctness of the assumptions we have made here, because we know the exact solution for the linear case.

As follows from the expression for the flux $J$, for a constant gradient $s = 1$ the system behaves as a thermal insulator, $J \sim N^{-3/2}$, and for $s = 0$ the flux does not depend on system size. In the intermediate case of $s = 2/3$, the system conforms to Fourier's law $J \sim N^{-1}$. Note that the dependence of heat flux on the size of a harmonic system is specified by boundary conditions \cite{determinant,Reich}, and it is possible to select them such that Fourier's law in the system will be obeyed. In the $s = 2/3$ case we selected appropriately the system itself while leaving unchanged  standard boundary conditions. Thus a system with linearly increasing masses can be adjusted such that it will conform to Fourier's law. 

Let us turn now to the temperature profile. First, as seen from relations \eqref{eq:green}, if the masses are not constant within the chain, the temperature at the points at which $M'(a)=0$ should rise strongly.

Second, it can be shown that in the case of constant masses thermal profile within the system $T(a) = T$ is constant, a point well enough known. In the reverse approximation, in the case of linearly increasing masses, when $M_1 \gg M_{0}$, we come to:
$$T_a=T- \frac{1}{\pi} \Delta T  \arcsin \left(\sqrt{a}\right).$$
We see immediately that the temperature in the system is determined by the more massive end. Now at the center of the system under consideration a temperature gradient $ - \Delta T/\pi$ will appear.

This conclusion correlates well with available numerical simulation \cite{PhysRevB.76.020301}. Temperature gradient is built up in the graded harmonic chain, but Fourier's law does not hold as mass difference does not grow with the system size $N$ as happens in the present work.

In conclusion, we have presented the solution to systems of linear equations \eqref{eq:x} describing systems of functional-gradient  materials. In particular, we analyzed the problem of a nonequilibrium steady state of a harmonic variable-mass system connected with heat reservoirs which are maintained at different temperatures.

We have shown that in the particular case of linearly increasing masses the heat flux depends on system size $J \sim 1/N^{3s/2}$, with the exponent in this relation being determined by the law governing the increase of the boundary mass $M \sim N^s$. This result finds ready explanation when we turn to Fig. \ref{fig:dispersion}. The systems under consideration maintain vibrations of two types, delocalized phonons which transport heat ballistically and localized ``gradons''. By properly varying the boundary mass as a function of $N$, we modulate in this way the number of phonons, to arrive finally at the dependence of the heat flux $J$ on the system size $N$. A similar effect accounts for the appearance of a thermal gradient in a system with linearly increasing masses. We have demonstrated in the particular example of a system with linearly increasing masses that a harmonic system can both sustain formation of a temperature gradient and conform to Fourier’s law.

\begin{acknowledgments}
I wish to express gratitude to A. Ya. Vul' for the assistance he has been rendering throughout my work on the paper, as well as to K. Yu. Platonov for our discussions which assisted in formulation of this problem. I would further like to thank A.M. Samsonov for useful discussions.  
\end{acknowledgments}

\end{document}